# Microstructural analysis and optical properties of $Cs_2BiAgBr_6$ halide double perovskite single crystals

O.A. Lozhkina[1*], A.A. Murashkina[1], M.S. Elizarov[1], V.V. Shilovskikh[1], A.A. Zolotarev[1], Yu.V. Kapitonov[1], R. Kevorkyants[1], A.V. Emeline[1], T. Miyasaka[2]

[1] Saint-Petersburg State University, ul. Ulyanovskaya 1, Saint-Petersburg, 198504, Russia.

[2] Toin University of Yokohama, and Peccell Technologies, Inc., 1614 Kurogane-cho, Aoba, Yokohama, Kanagawa 225-8502, Japan.

[*] Corresponding author.

Corresponding author's     e-mail address:     st040462@student.spbu.ru

postal address:     Saint-Petersburg State University, ul. Ulyanovskaya 1, Saint-Petersburg, 198504, Russia.

**Abstract**

Single crystals of $Cs_2BiAgBr_6$ lead-free perovskite were grown by crystallization from supersaturated solution. According to characterization by XRD and EBSD methods the double perovskite single crystal are of the cubic *Fm*-3*m* symmetry with the lattice constant $a$ = 11.20 Å. DFT predictions based on the single crystal X-ray diffraction analysis reveal that the material is an indirect band gap semiconductor. Low temperature (1.4 K) photoluminescence spectra demonstrate three broadened bands that correspond to two lowest computed indirect and one direct band-to-band transitions.

# 1. Introduction

During the last few years lead-based halide perovskites $APbX_3$ have attracted thorough attention of alternative energy research community due to their very efficient conversion of solar energy into electric current [1]. Further studies showed that these materials are also applicable for a design of optoelectronic devices including lasers and light-emitting diodes [2], photo- and γ-ray detectors [3-5] etc. One of the characteristics of these ionic crystals is a preservation of their important optical properties even in the case of low-cost wet chemistry synthesis. This unique feature is a direct sequence of their defect tolerant electronic structure [6].

Along with useful features, lead-halide perovskites demonstrate several drawbacks. Among them is low stability with respect to humidity and temperature, and high toxicity due to the content of lead. These drawbacks restrict potential use of devices based on these materials. For this reason lead-free perovskites with optoelectronic properties similar to those of $APbX_3$ are extensively sought nowadays. However, it should be kept in mind that 5s electron pair of lead is responsible for the formation of defect-tolerant band structure of lead-halide perovskites [6]. Following these lines, perovskites based on the electronic analogs of lead, namely, tin and germanium, were synthesized. Unfortunately, these materials demonstrate significant instability with respect to oxidation [7]. Another possible way to avoid usage of lead is to substitute it by two heterovalent atoms $B'^{3+}$ and $B''^{1+}$ thereby forming the so-called double perovskites, which are ionic solids with the chemical formula $A_2B'B''X_6$, where $A^+$ is organic or inorganic cation; $B'^{3+}$ can stand for $Bi^{3+}$ or $Sb^{3+}$ that possess valent $ns^2$ lone pair and are known to be of a good chemical stability; $B''^+$ can stand for $Ag^+$, $Cu^+$, $Au^+$, $In^+$ or other monovalent cation; and $X^-$ represents a halogen anion. The structures of conventional $ABX_3$ and double $A_2B'B''X_6$ perovskites are depicted in the Figure 1. Research in this direction has revealed some difficulties in the synthesis of these compounds. For instance, $In^+$-containing double perovskites tend to oxidize [8], $Cu^+$ cations form no hexacoordinated complexes [9], and $Au^+$-containing substances undergo photoreduction [10].

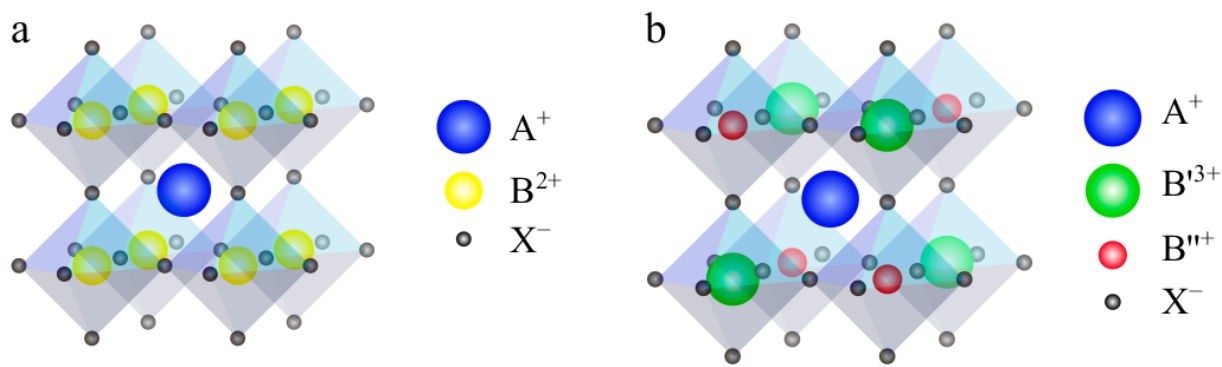

Figure 1.     The structures of conventional $ABX_3$ (a) and double $A_2B'B''X_6$ (b) perovskites.

The double perovskite $Cs_2SbAgBr_6$ should be unstable due to the small ionic radius of $Sb^{3+}$. However, this cation can be used to dope considerably more stable compound $Cs_2BiAgBr_6$, doping with the $Sb^{3+}$ cation paves the way to tuning of the perovskite's electronic band gap. It was found that inclusion of $Sb^{3+}$ narrows it[11]. Width of a band gap in double perovskites can also be tuned by exchanging halogen anions. In was showed that the band gap of $Cs_2InAgCl_6$ perovskite can be narrowed by partial substitution of $Cl^-$ anions with $Br^-$ [12]. Other two double perovskites, $Cs_2BiAgCl_6$ and $Cs_2BiAgBr_6$, synthesized recently, are known to be highly resistive to the presence of water and oxygen in the air [13]. This, together with the fact that the widths of their electronic band gaps are suitable for absorption of visible light, makes them good alternative to lead-halide perovskites.

Single crystals are known to be the best suited for fundamental studies of optical properties of crystalline compounds. In this work, the solution processed single crystals of $Cs_2BiAgBr_6$ underwent microstructural analysis followed by their optical characterization. Due to the recent interest in perovskite single crystals and, in certain cases, epitaxial heterostructures based on halide perovskites [14-17], at present, methods for local structural analysis are in high demand. Here, we discuss those of them, which are based on Scanning Electron Microscopy (SEM).

## 2.     Crystal growth

Single crystals of $Cs_2BiAgBr_6$ double perovskite were synthesized by crystallization from supersaturated solution [18]. For this purpose, first, the acidic solution of 0.1 mol/l $BiBr_3$, 0.1 mol/l AgBr, and 0.2 mol/l CsBr in 50% HBr was heated to 150 ºC and subsequently cooled down to room temperature at the speed of 10 ºC/h. The precipitated dark-purple crystals (Fig. 2a) were filtered and air-dried. The crystals look transparent and are octahedral in shape. The latter is typical of a cubic crystal lattice. The surface macro-defects seen in reflected light (Fig. 2b) are formed due to evaporation of solvent remained on the crystals after filtration.

3.     **Structural characterization**

X-ray diffraction measurements were performed using Oxford Diffraction single crystal diffractometer "Xcalibur". According to X-ray analysis the synthesized crystals possess cubic lattice symmetry *Fm*-3*m* with the lattice constant $a$ = 11.20 Å. The crystal's microstructure was analyzed with Hitachi S-3400N SEM, equipped with Oxford Instruments X-Max 20 extension for Energy-Dispersive X-ray spectroscopy (EDX) and AZtecHKL Channel 5 extension for Electron Backscatter Diffraction (EBSD). The fact that the obtained samples consist of one crystalline phase is confirmed by SEM images, which are produced in Backscattered Electron (BSE) regime sensitive to phase contrast (Fig. 2c). In Secondary Electron (SE) regime, which is sensitive to the surface relief, one can see small macro-defects (Fig. 2d) whose origin is described in the previous section. Stoichiometry of the samples analyzed by EDX (Fig. 3a) matches the formula $Cs_2BiAgBr_6$.

EBSD is a powerful tool for local structural analysis and determination of crystallographic orientation of samples with small spatial resolution. The latter is determined as a contact area of an electron beam with sample's surface. For the electron beam of 10 keV it scales to a few microns. The Figure 3b shows EBSD pattern of the synthesized $Cs_2BiAgBr_6$ single crystal together with the fit based on XRD analysis. The fit of Kikuchi lines allows for determination of the crystal's orientation (Figure 3; inset b). Finally, we would like to note that in contrast to organic-inorganic perovskites the crystals of $Cs_2BiAgBr_6$ are resistant to electron beam [19].

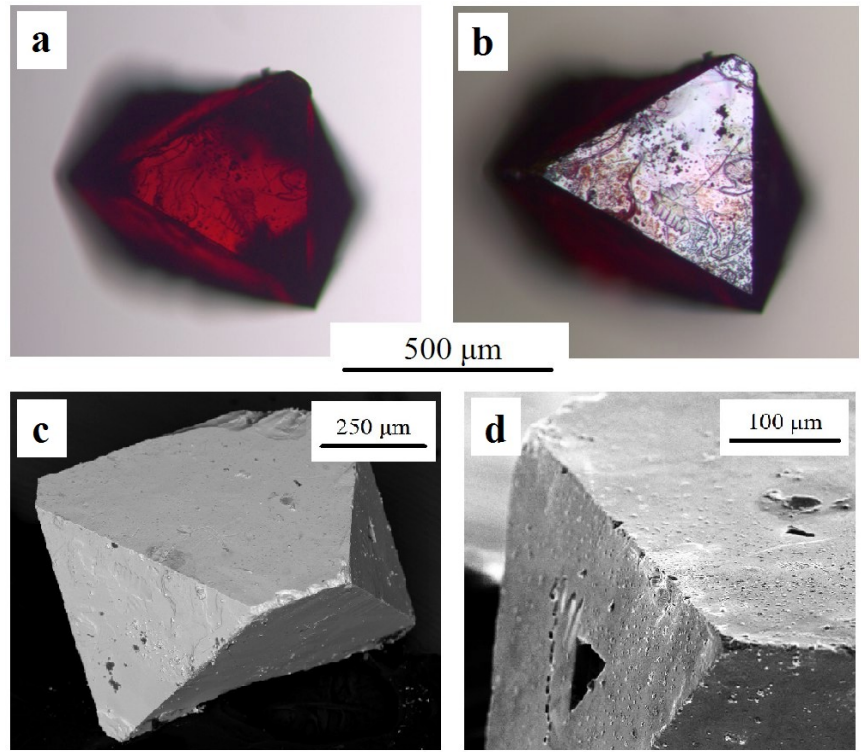

Figure 2. Optical microphotographs in transmitted (a) and reflected (b) light and the SEM images in BSE (c) and SE (d) regimes of the single crystals of $Cs_2BiAgBr_6$.

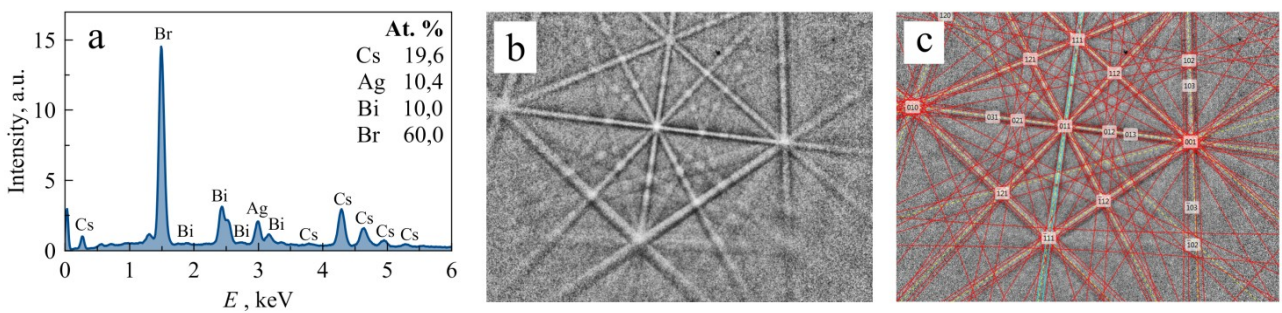

Figure 3. EDX-spectrum (a) and EBSD diffraction pattern (b) with fit for the single crystals of $Cs_2BiAgBr_6$ (c).

## 4. Method of Computation

Band structure (BS) of the double perovskite $Cs_2BiAgBr_6$ was modeled in the framework of periodic Density Functional Theory using Abinit-8.2.3 program [20]. LDA density functional [21] was applied. Spin-orbit coupling was accounted for by using relativistic separable Hartwigsen-Goedecker-

Hutter pseudopotentials [22]. In the computations, Brillouin Zone was sampled over 8x8x8 Γ-centered Monkhorst-Pack grid [23]. The applied kinetic energy cut-off was equal to 100 Hartree. The crystal lattice structure of $Cs_2BiAgBr_6$ perovskite corresponds to the spatial symmetry group *Fm-3m*. BS was computed over 61 points along the high symmetry path L-Γ-X-W.

5. **Experimental setup of photoluminescence measurements**

For the photoluminescence (PL) study the single crystal of $Cs_2BiAgBr_6$ was placed into closed loop helium cryostat and cooled down to 1.4 K. The pumping was done using cw-laser with the wavelength of 450 nm. The beam's diameter on a surface of the crystal was around 100 μm. The recorded PL spectrum is depicted in the Figure 4b.

6. **Optical transitions: results and discussion**

The Figure 4a depicts the computed BS of $Cs_2BiAgBr_6$. Clearly, the material is indirect band gap semiconductor. CB minimum of the BS is located at the point Γ of Brillouin Zone (BZ). VB demonstrates two energetically degenerate maxima at the points L и W. The second VB maximum occurs at the point k0, which is at a half distance from Γ to L i.e. at [0.25; 0.25; 0.25] while the third one is at the Γ-point of VB. Thus, the two degenerate energy lowest indirect transitions are from the point Γ (CB) to the points L (VB) and X (VB). These are followed by the indirect transition from the point Γ (CB) to the point k0= [0.25; 0.25; 0.25] (VB) and the direct transition at Γ-point. The next series of possible indirect transitions is more than 0.3 eV higher in energy.
The measured quantum yield of PL of the $Cs_2BiAgBr_6$ single crystals is several magnitudes lower than that of its direct band gap lead-based analogue $CsPbBr_3$. This is in agreement with our DFT calculations predicting indirect nature of a band gap of the synthesized compound. In the Table 1 we compare computed DFT transitions with those obtained from the PL experiments. The most intense and energetically lowest is the PL peak centered at 1.946 eV. This energy corresponds to the computed indirect Γ-L and Γ-X transitions. The two more peaks of lesser intensity at 2.095 eV and

2.254 eV we attribute to the indirect DFT Γ-k0 and to the direct DFT Γ-Γ transitions, respectively. No higher energy optical transitions were observed in the PL spectra. Thus, the number of observed optical transitions matches that of the first transition series computed with DFT.

Table 1. Computed (DFT) and experimental (PL) opticl transions in $Cs_2BiAgBr_6$.

|  | DFT modelling | | PL spectra | |
| --- | --- | --- | --- | --- |
|  | E, eV | ΔE, eV | E, eV | ΔE, eV |
| Γ-L and Γ-X | 1.782 | 0 | 1.946 | 0 |
| Γ-k0 | 1.820 | 0.038 | 2.095 | 0.149 |
| Γ-Γ | 1.899 | 0.117 | 2.254 | 0.308 |

Half width at half maxim (HWHM) of the first and the second peaks is equal to ~70 meV, while that for the third maximum is nearly twice larger. These large HWHM values observed at liquid-helium temperatures when thermic homogeneous broadening is negligible as well as the Gaussian shape of bands imply considerable inhomogeneous broadening of the PL bands.

Such a significant inhomogeneous broadening of the luminescence bands indicates the absence of the defect tolerance of this structure. Thus, substitution of lead in the lead-halide perovskites with two heterovalent cations if only one of those possess lone-pair on valent s-orbital does not allow preserving the outstanding optical properties of the material.

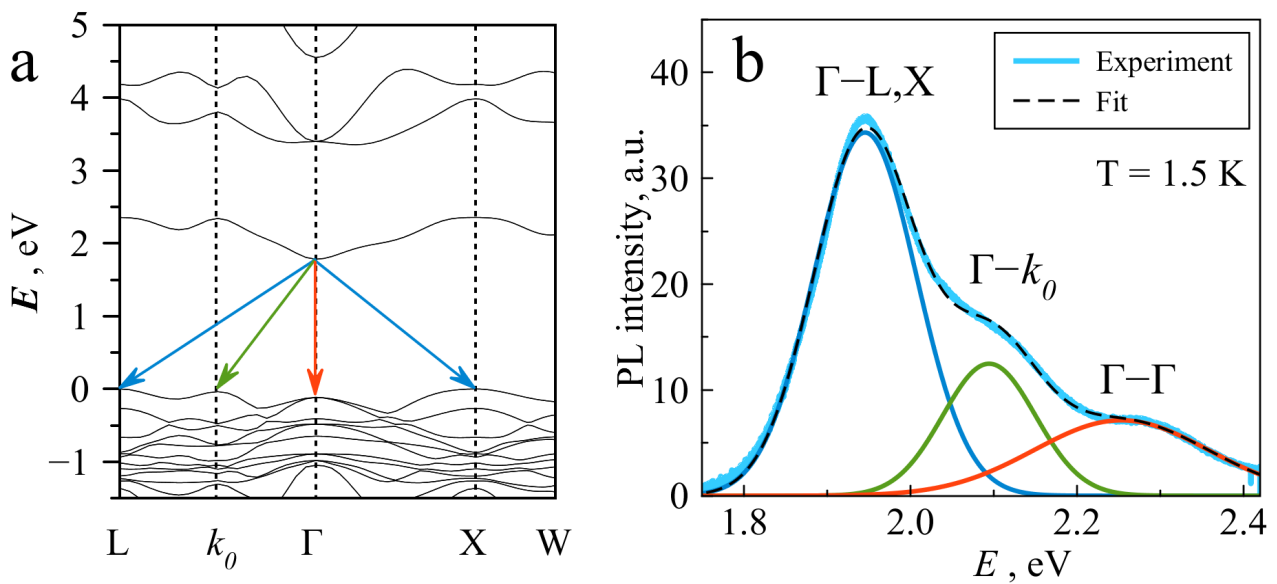

Figure 4. DFT band structure of $Cs_2BiAgBr_6$ (a) and low temperature PL spectrum of the $Cs_2BiAgBr_6$ single crystals (b): blue solid line – raw PL data, black dashed line – approximation by three Gaussian functions.

## 7. Conclusions

In the present study single crystals of $Cs_2BiAgBr_6$ double perovskite were synthesized by crystallization from supersaturated solution. Single crystals are formed in cubic lattice *Fm*-3*m* symmetry with the lattice constant *a* = 11.20 Å as demonstrated by XRD and EBSD methods. Low temperature luminescence exhibits three bands with maxima 1.946 eV, 2.095 eV, and 2.254 eV, which according DFT modeling of the band structure correspond to first two indirect and one direct transitions. The complex band structure and indirect optical transitions in double $Cs_2BiAgBr_6$ perovskite result in low quantum yield of photoluminescence comparing to the lead perovskites.

**Acknowledgments**

The present study was performed within the Project "Establishment of the Laboratory "Photoactive Nanocomposite Materials" No. 14.Z50.31.0016 supported by a Mega-grant of the Government of the Russian Federation. A.A. Murashkina and A.V. Emeline are grateful to RFBR 17-53-50083 JF_a grant which supported the studies concerning the synthesis and photoluminescence measurements. This work was carried out using equipment of SPbU resource centers "Nanophotonics", "Geomodel", 'Computational Center', and "X-ray Diffraction Studies".